\documentclass[reprint,secnumarabic,longbibliography,amssymb,aps,prd,showpacs,floatfix,nofootinbib]{revtex4-2}
\usepackage[utf8]{inputenc}
\usepackage{graphicx}
\usepackage[squaren]{siunits}
\usepackage{amsmath,amsfonts,amssymb}
\usepackage{siunits}
\usepackage{xcolor}

 \usepackage{verbatim}
\usepackage{multirow}
\graphicspath{
   {../figures/}
}

\begin{document}

\title{Unfocused laser ignition of high-pressure He--H$_2$-O$_2$ combustible mixtures}%

\author{R. \surname{Grosso Ferreira}$^1$}%
%\email[]{ricky_charizard@hotmail.com}
\author{B. \surname{Carvalho}$^1$}%
%\email[]{bernardo.carvalho@tecnico.ulisboa.pt}
\author{J. \surname{Rodrigues}$^{1,2}$}%
%\email[]{jose.rodrigues@tecnico.ulisboa.pt}
\author{R. \surname{Rodrigues}$^1$}%
%\email[]{rafael.rodrigues@tecnico.ulisboa.pt}
\author{A. \surname{Smith}$^3$}%
%\email[]{arthur.smith@fluidgravity.co.uk}
\author{L. \surname{Marraffa}$^4$}%
%\email[]{lionel.marraffa@esa.int}
\author{M. \surname{Lino da Silva}$^1$}%
%\email[]{mlinodasilva@tecnico.ulisboa.pt}

\affiliation{$^1$Instituto de Plasmas e Fus\~{a}o Nuclear, Instituto Superior T\'{e}cnico, Universidade de Lisboa, Av. Rovisco Pais, 1049--001, Lisboa, Portugal}

\affiliation{$^2$Departamento de Física, Faculdade de Ciências e Tecnologia, Universidade do Algarve, Campus de Gambelas, 8005-139, Faro, Portugal}

\affiliation{$^3$Fluid Gravity Eng.\protect, The Old Coach House, West Street, Emsworth, Hampshire, PO10 7DX, United Kingdom}

\affiliation{$^4$European Space Agency/European Space Research and Technology Centre, Aerothermodynamics Section keplerlaan 1, PO Box 299, 2200AG, Noordwijk, The Netherlands}

\date{\today}%

\begin{abstract} %Popiular summary 250 words max. Current word count: 191 words

%Laser irradiation provides a clean, non-local, non-intrusive and low quenching alternative to spark and hotwire systems for the ignition of combustible mixtures.

We report consistent ignition of high-pressure ($p>20-30$~\bbar) hydrogen-oxygen mixtures diluted with helium, using an unfocused Nd:YAG laser. This corresponds to laser irradiances several orders of magnitude below the minimum ignition energies reported in the literature. %We investigate the optical absorption of the laser signal in non-combustible hydrogen-helium and oxygen-helium mixtures at pressures up to 100~\bbar.

By placing a mirror inside a cylindrical vessel and filling it up to 100~\bbar\ with H$_2$--He or O$_2$--He non-combustible mixtures, we obtain the pressure-dependent absorptivity of the combustible He--H$_2$--O$_2$ mixture. We find no measurable absorption of the laser signal by the medium, for the overall pressure range, to the experimental apparatus sensitivity (about 1\% of the laser irradiance). This leads credence to the theory that ignition stems from seed electrons created by autofocusing ionization of dust/impurities in the gas.

\end{abstract}

%\pacs{82.33.Vx, 42.55.Rz, 51.70.+f}
\maketitle
% \tableofcontents

%\section{Coisas Importantes}
%
%
%
% 
%
%
%
%\begin{table}[htbp]
%\centering
%\caption{ESTHER laser ignition system parameters}
%\label{Tab: Laser-Parameters}
%\begin{tabular}{l|r l}
%$E_{pulse}$ (mJ) 		& 193 &$\pm$ 2 \\
%\hline
%$\Delta t_{pulse}$ (ns)	& 5 &	\\
%\hline
%$A_{beam}$ (mm$^2$) 	& 23.15 &$\pm$ 0.43 \\
%\hline
%$\mathcal{T}_{window}$ 	& 0.801 &$\pm$ 0.023 \\
%\hline
%$I$ ($10^8$  W/cm$^2$) 	& 1.336 &$\pm$ 0.115	 
%\end{tabular}
%\end{table}

\section{Introduction}

Laser-induced spark ignition has experienced increasing interest in recent years due to its advantages over conventional spark-plug ignition. It is specially effective in high pressure regimes where conventional spark plugs have low life time \cite{Weinrotter-2004, Kopecek-2003}. Ignition occurs when a 10$^{-3}$ fraction of the gas is ionized \cite{Ref-46-Phuoc-2006}. Seed electrons are acelerated by the laser's energy, collide with other molecules, ionizing them, creating an electron avalanche, and leading to gas breakdown \cite{Phuoc-2006, Bradley-2004, Srivastava-2009}. The defining parameter that ensures ignition is the laser's irradiance $I$ (power per unit area). This is defined in Eq. \ref{Eq: Irradiance}, where $E_{pulse}$, $\Delta t_{pulse}$ and $w$ are the pulse energy, pulse time width and beam radii at focal point, respectively. Several authors studied laser-induced spark ignition in different conditions and gas mixtures, and reported the minimum irradiance values for initiating combustion. These are summarized in Table \ref{Tab: MIE-Literature}.   

\begin{equation}
I = \frac{E_{pulse}}{\Delta t_{pulse}} \; \frac{1}{\pi w^2 } \; .
\label{Eq: Irradiance}
\end{equation}

\begin{table}[htbp]
\centering
\caption{Minimum breakdown or ignition irradiance reported in the literature.}
\label{Tab: MIE-Literature}
\begin{tabular}{c|c|c}
$I$(W/cm$^2$) 			& Reference 							& Conditions/Model 				\\
\hline
1.93$\times 10^{11}$  	& Bradley, \cite{Bradley-2004} 		& Air at 1~MPa \\
1$\times 10^{10}$ 		& Phuoc, \cite{Phuoc-2006} 			& Theoretical \\
1$\times 10^{12}$ 		& Phuoc, \cite{Phuoc-2000}			& \begin{tabular}[c]{@{}c@{}}Air, CH$_4$, O$_2$, N$_2$,\\ H$_2$ at 150 and 3040~Torr\end{tabular} \\
1$\times 10^{14}$ 		& Srivastava, \cite{Srivastava-2009} 	& Drude model \\
1$\times 10^{12}$ 		& Weinrotter, \cite{Weinrotter-2004} 	& H$_2$-air 2.8~MPa air:fuel ratio 3 \\
1$\times 10^{10}$ 		& Lee, \cite{Lee-2001} 				& Propane-air at 1 atm
\end{tabular}
\end{table}

\begin{figure}[htbp]% order of placement preference: here, top, bottom
\centering
 \includegraphics[width=0.5\textwidth]{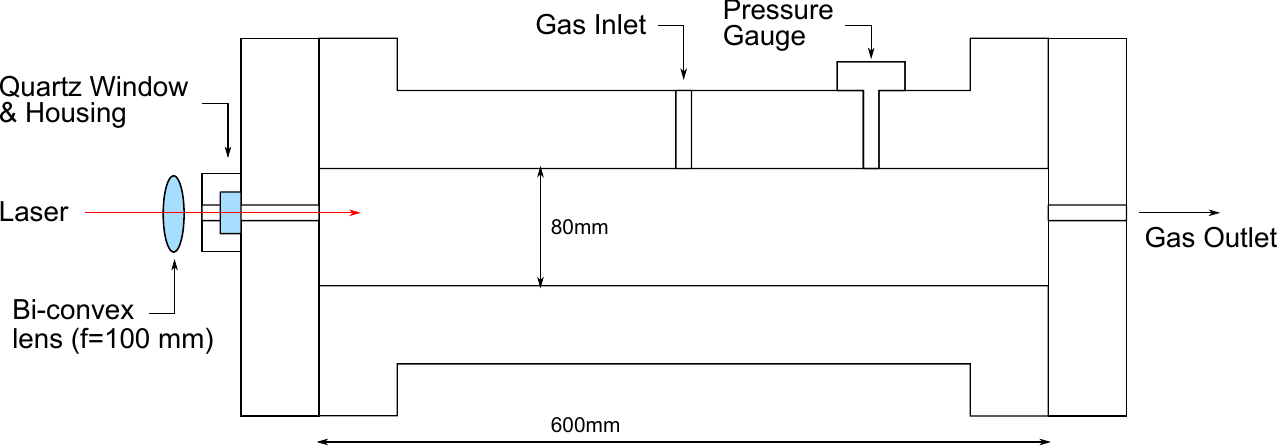}
 \caption{Schematic cutaway of the ``bombe'' and laser setup}
 \label{fig:bombe-scheme}
\end{figure}

The European Shock Tube for High Enthalpy Research (ESTHER) \citep{Mario-Lino-Esther, ESTHER-Qualification} is a facility capable of reaching shock-speeds in excess of 10 km/s. The facility comprises a 50.3 liter combustion chamber driver where an hydrogen/oxygen mixture diluted in helium with filling pressures up to 100~\bbar\ is ignited to a post-combustion pressure up to 600~\bbar. A test-scale model (3 liter) ``bombe" was firstly built to demonstrate the full driver and to ensure the safety and validation of all subsystems. The initial configuration deployed a hotwire ignition system which was later advantageously replaced by a laser ignition system. The resulting setup is depicted in Fig. \ref{fig:bombe-scheme}. During the test campaign campaign it was found that for filling pressures above 20-30 \bbar\ the laser could ignite the gas mixture without the use of the lens, reducing the irradiance to around 10$^8$ W/cm$^2$. This value is two orders of magnitude below the reported minimums in the literature, which always consider a focusing lens in the setup. 

\begin{figure}[htbp]
\centering
\includegraphics[width= 0.48 \textwidth]{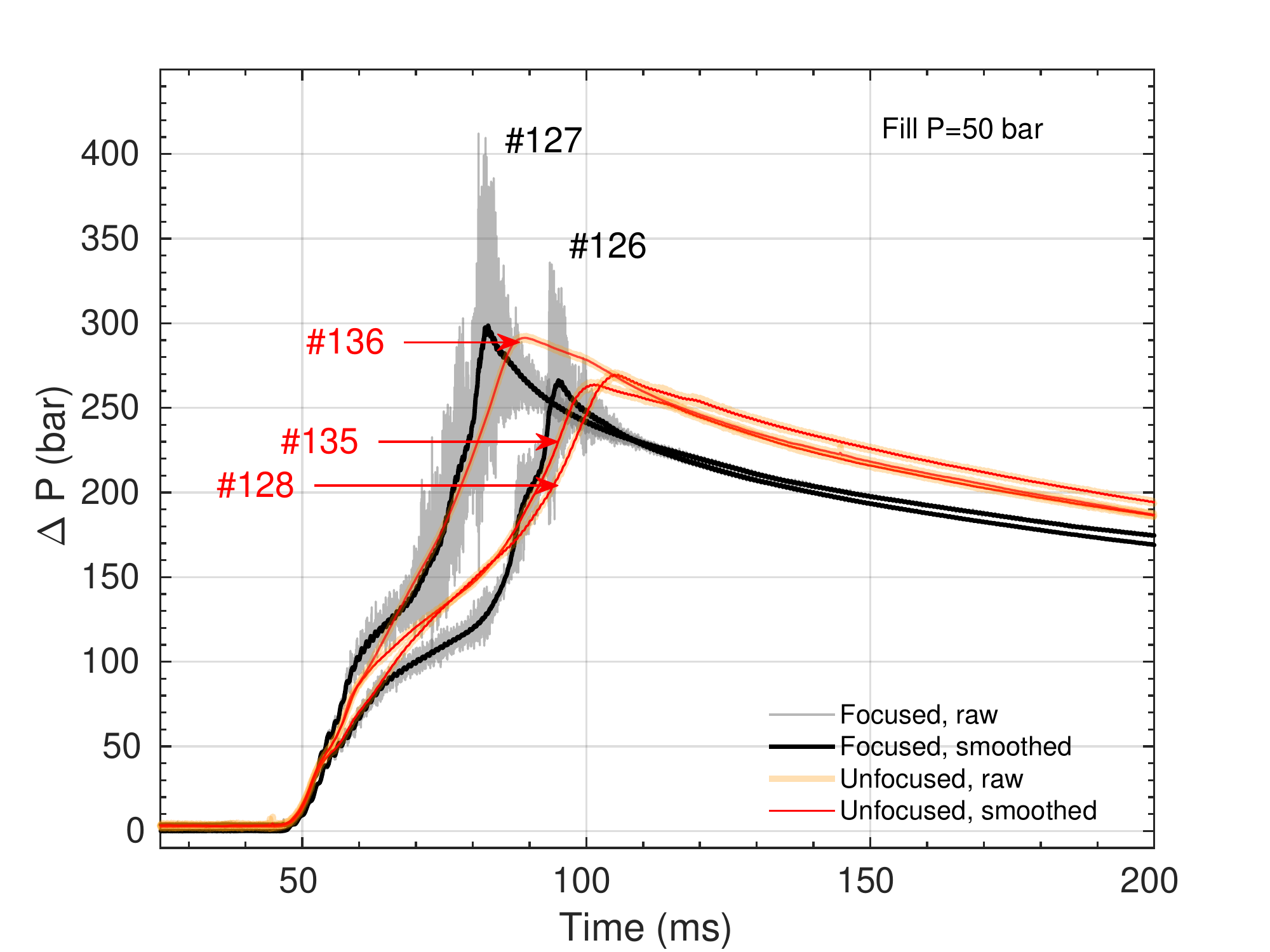}
\caption{Pressure rise $vs.$ time of the combustion chamber ignited with both focused and non-focused laser beam.}
\label{fig:bombe-deflagration}
\end{figure}

Figure \ref{fig:bombe-deflagration} presents the recorded pressure signal in both focused and unfocused ignition mode. A series of shots (\#126, \#127, \#128, \#135, \#136) were measured for a filing pressure of 50~\bbar. The helium dilution is around 70~\% in terms of molar fractions, and the mixtures range from rich (\#127 and \#126 with $\varphi=1.2$ and $\varphi=1.1$ respectively), stochiometric (\#127 with $\varphi=1.0$), to poor (\#128 and \#135 with $\varphi=0.8$ and $\varphi=0.7$ respectively). Shots \#126 and \#127 were carried with a focusing lens, the others with the laser remaining unfocused.

One may further note that the combustion dynamics change significatively depending on whether the laser is focused or not. The focused shot pressure signals are considerably noisy, hinting at localized transition to detonation during the pressure rise, whereas the unfocused shots evidence smooth subsonic deflagrations. These and other phenomena (such as the shape of the pressure ride) will not be discussed in detail here, and will be left for another upcoming publication by our group. The objective of this study is instead to report laser absorption characteristics in high pressure He--H$_2$--O$_2$ environments and infer on the possible mechanisms behind laser-induced spark ignition.

%%%%%%%%%%%%%%%%%%%%%%%%%%%%%%%%%%%%%%%%%%%%%%%%

\section{Experimental Setup}
\label{Sec: Experiment}

The experimental setup comprises two main elements: a high-pressure combustion chamber a high-power pulsed laser. The combustion chamber has an associated gas filling system and the laser has an associated beam conditioning system. A cylinder was positioned inside the combustion chamber, at the opposite side of the window to support a 0º mirror used to reflect the high power beam that enters the window. The window itself is a fused silica cylinder of 50.8 mm diameter and 10 mm thickness. 

\begin{figure}[htbp]
\centering
\includegraphics[width=0.49\textwidth]{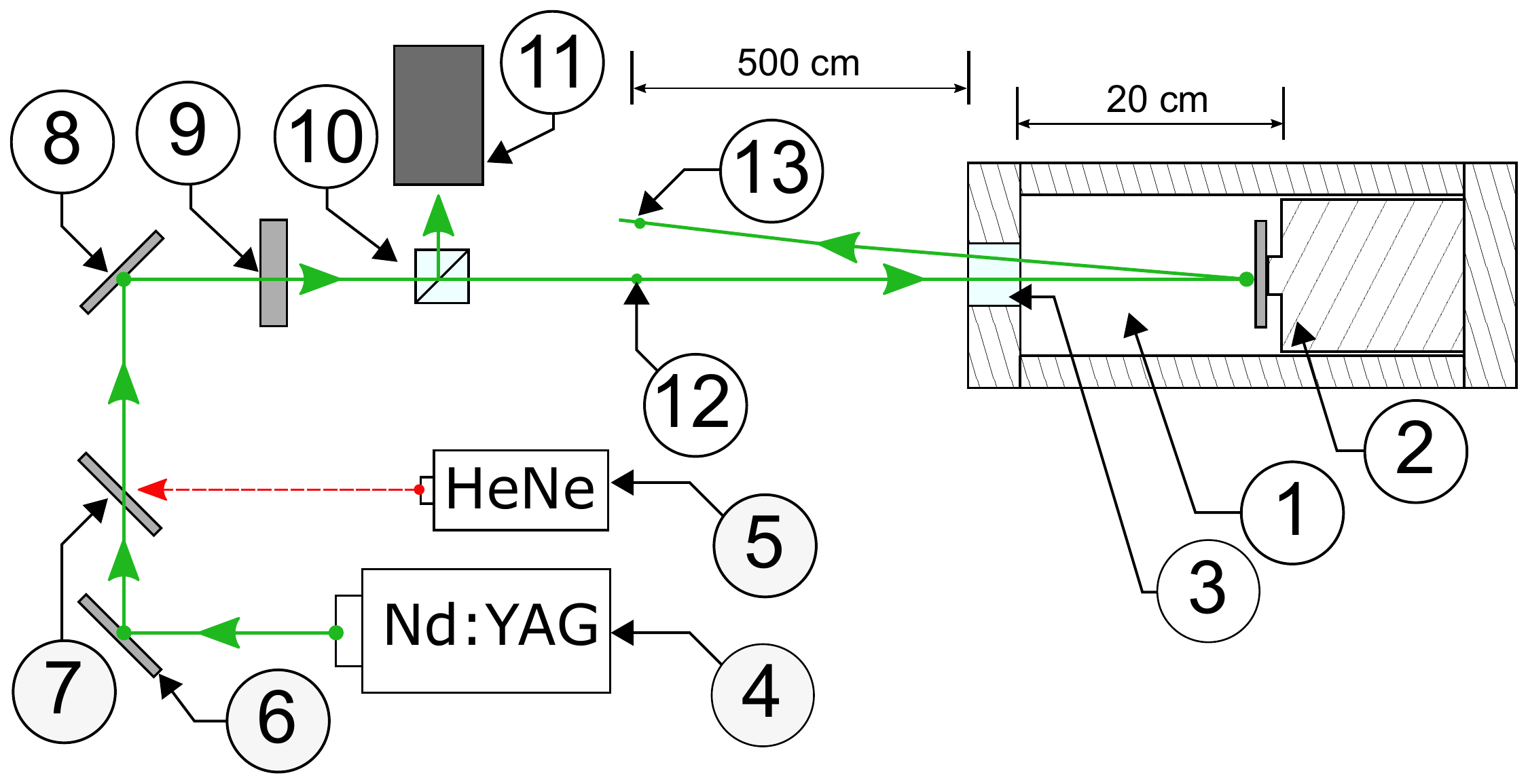}
\caption{Schematic for the Laser absorption experiment, not to scale. 
1, 2, 3 - elements in the ``bombe";
4-13 - elements on the optical breadboard.\\
1 - Combustion chamber; 
2 - chamber reduction cylinder and 0º mirror;
3 - Fused silica window    
4 - HeNe laser;
5 - Nd:YAG laser ($\lambda_0$= 1064 nm); 
6 - 45º mirror;
7 - Beam combiner;
8 - 45º miror;
9 - Half-wave plate; 
10 - Beam splitter cube;
11 - Beam dumper; 
12 and 13 - Laser power reading location before and after crossing the ``bombe".}
\label{fig:laser-setup}
\end{figure}

The gas filling system description may be found in \citep{BBC-2016}. The high-power laser and the beam conditioning elements were positioned on an optical breadboard, in front of the optical port of the ``bombe". The high-power laser was a Quantel Brilliant laser, a 10~Hz Q-switched Nd:YAG laser emitting 200~mJ pulses at 1064~nm, its fundamental wavelength. The laser was placed facing the opposite of the line of sight towards the ``bombe" window, with two 45º mirrors, allowing height and azimutal deviation, deflecting the laser beam towards the ``bombe". To help the alignment, a CW He-Ne laser beam with 3~mW at 632.8~nm was combined with the Nd:YAG one. In such way, and for security reasons, the Nd:YAG laser was kept switched off whenever possible. The half-wave waveplate and the beam-splitter were used to regulate the beam power, a common procedure with linearly polarized laser beams, as is the case. The unwanted power was diverted from the beam splitter cube to the beam dump. The mirrors, beam spitter, beam combiner and half-wave beamplate were from CVI. The only two optical elements not installed in the optical breadboard were the silica window in the optical port of the ``bombe" and the 0º mirror in the end of the combustion chamber. This last mirror was reflecting at an angle slightly off-axis, to separate the entering beam from the exiting one, allowing the measurement of the laser beam power in both situations on the breadboard by a powermeter, a Coherent Molectron Powermax 500A. The schematics of the setup is presented in Fig. \ref{fig:laser-setup}.

\subsection{Experimental procedure}

%The experiment follows these steps:

The laser was run at 10 Hz, the beam entering the combustion chamber with pulses averaging 200 mJ after conditioning. No focusing of the laser occurs throughout its optical path. The chamber is filled with an He--O$_2$ [10:1] and then an He--H$_2$ [9:2] mixture up to a pressure of 100 bar. This mimics a nominal [8:2:1] He--H$_2$--O$_2$ combustible mixture without the associated reactivity. Repeating the experiment twice with those two mixtures allows pinpointing the influence of chemical composition from any differences in absorption that may arise between these two sets of experiments. Then, the absorptivity of the He--H$_2$--O$_2$ mixture may be correlated from both partial mixtures absorptions.

The initial measurement is carried out at maximum laser irradiance, with subsequent measurements being carried out at lower irradiances, adjusted using the half-wave plate. After about 5 to 10 different input powers, the pressure in the chamber is reduced using the venting valve, and the procedure is repeated. Once the first mixture (He--O$_2$ [10:1]) is concluded, a second run is carried out in a similar fashion, considering a (He--H$_2$ [9:2]) mixture. %A difficulty encountered during this round of experiments was that the O$_2$ gas bottle did not have sufficient pressure to allow filling the combustion chamber to 100~\bbar, hence measurements for the He--O$_2$ mixture were carried out a maximum pressure p=65~\bbar, unlike measurements for the He--H$_2$ mixture which proceeded up to 100~\bbar.

To make a final cross-check of the experiment, allowing us to evaluate any spurious power losses through air and optical components, another run of the experiment was made with an open chamber and without the silica window, and a final run measuring the power loss between points 12 ($P_{in}$) and 13 ($P_{out}$). No measurable loss to the sensitivity of our powermeter could be identified.

%A limitation of our experiment is the poor transmittance of the fused silica window which has to have a very large thickness to withstand operational pressures up to 600~\bbar\ in the combustion vessel. The transmittance of the window was about 80\% in the laser 1064~\nano\metre\ wavelength. Furthermore, it was found out that inhomogeneities inside the glass also led to scattering of radiation inside, which means that an additional amount of radiation will be lost as a result of having part of the beam deviated and hitting elsewhere than the mirror at the end of the combustion chamber. 
A measurement of input/output power was carried out with the window and the mirror placed inside the combustion chamber in room air, yielding an ``effective'' transmittance of about 55\%

\section{Results and Discussion}
\label{Sec: Results}

\begin{figure}[htbp]
\centering
\includegraphics[width=0.45\textwidth]{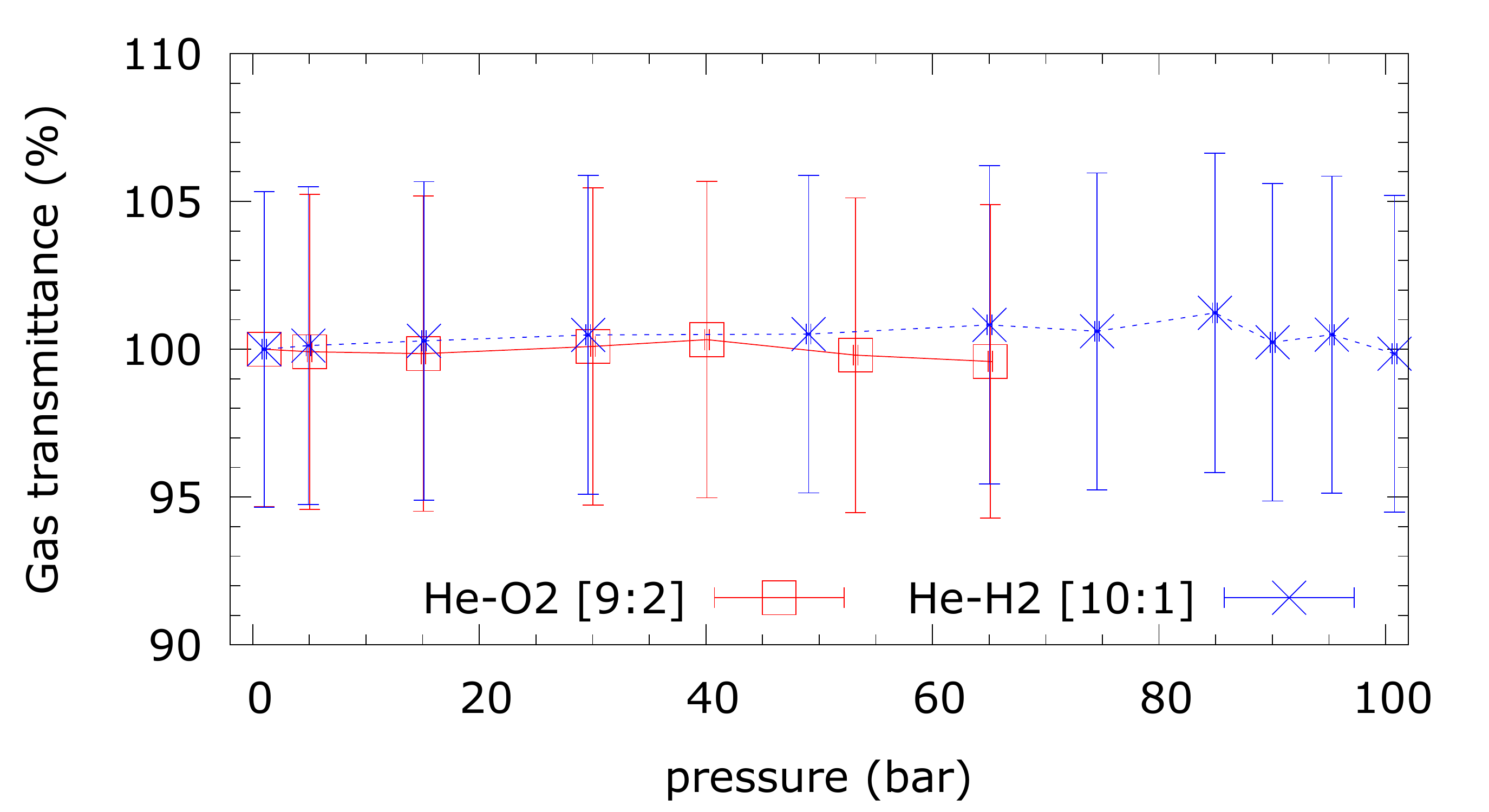}
\caption{Test gas transmittance as a function of filling pressure, for mixtures of He--O$_2$ [10:1] and He--H$_2$ [9:2]. Error bars above one hundred have no physical meaning}
\label{fig:transmittance}
\end{figure}

The transmittance $\mathcal{T}$ of the setup was calculated using Beer-Lambert's law, see eq. \ref{Eq: Beer-Lambert}, where $P_t$ and $P_0$ are the transmitted and original beam power. The values are then corrected by the silica window transmittance, to have the gas transmittance values, presented in Fig. \ref{fig:transmittance}. Gas transmittance is constant and around 1 for the pressure range between 1 and 100 bar, and for both mixtures. This means the gas is optically transparent to the 1064 nm laser radiation, even though at 100 bar there is 100 times more matter inside the chamber.

\begin{equation}
P_t / P_0 = \mathcal{T} . %= \exp( -\alpha_{abs} d) .
\label{Eq: Beer-Lambert}
\end{equation}

%\section{Discussion}
%\label{Sec: Discussion}

The experiments show that the laser energy is not absorbed directly by the gas mixture. The result is somewhat expected as there are no electronic transitions for helium, oxygen or hydrogen near the 1064 nm region to absorb the laser photons. The gas transparency to the laser implies that seed electrons are not generated by multiphoton ionization. In multiphoton ionization an atom/molecule absorbs a number of photons so that it releases an electron as it ionizes. Srivastava \cite{Srivastava-2009} states that to generate the first electrons the irradiance should be in 10$^{14}$ W/cm$^2$. Ionization energy of O$_2$ molecules is 12.07 eV, whereas a typical Nd:YAG laser at 1064 nm has a photon energy of 1.16 eV. Thus, 10 ou more photons are need to produce one free electron. Experiments and studies done by many different authors report different values for breakdown intensity, yet always with orders of magnitude below 10$^{14}$ W/cm$^2$. For example, Srivastava reports the intensity in the locus to be of 10$^{12}$ W/cm$^2$, while Phuoc in \cite{Phuoc-2006} even states 10$^{11}$ W/cm$^2$ to be suffcient. Several authors, \cite{Weinrotter-2004,Tauer-2007,Kopecek-2003,Kopecek-Reider-2003,Srivastava-2009}, defend that these seed electrons do not come from multiphoton ionization but from impurities in gas mixture (e.g. dust, aerosol or soot particles). This hypothesis is supported by a non-dependence of the minimum pulse energy (MPE) with the laser wavelength, and a strong pressure dependence reported in \cite{Kopecek-2003,Phuoc-White,Kopecek-Reider-2003,Phuoc-2000,Lee-2001}, whereas multiphoton ionization predicts a very weak dependence. The strong pressure dependence is evident in this work, where ignition could be achieved with a non-focused laser, but only if sufficiently high pressures are achieved (with ignition being achieved with multiple pulses above p=~20~\bbar, and with a single pulse above p=~30~\bbar. 

The laser has a pulse energy of 193$\pm$2 mJ, a time width of 5 ns and a measured area of 23.15$\pm$0.43 mm$^2$, which yields an irradiance of (1.336$\pm$0.115)$\times 10^8$ W/cm$^2$. This irradiance is not sufficient to ensure a visible electrical breakdown of room air, yet it is sufficient to ignite our high pressure combustible mixture. Phuoc in \cite{Phuoc-2006} asserts that once electrical breakdown is achieved, gas ignition will necessarily follow up. Nevertheless, it seems to be the case that ignition may be achieved without a macroscopic electrical breakdown of the gas 
{in the usual fashion, with a bright flash and the emission of noise. Nevertheless, it cannot be ruled out that an ignition kernel may develop as the result of gas breakdown owing to electron release by dust/impurities, however such phenomena is not visible our audible}

%\textcolor{blue}{It may be possible that electrical breakdown still occurs in the non-focused ignition mode, although at a microscopic scale. The dust/microparticles may absorb the laser energy and create the first seed electrons leading to the electron-cascade process, however the energy density is lower thus creating a smaller spark.} %This is what we report in this work.  

\begin{comment}

\begin{table}[htbp]
\centering
\caption{ESTHER laser ignition system parameters}
\label{Tab: Laser-Parameters}
\begin{tabular}{l|r l}
$E_{pulse}$ (mJ) 		& 193 &$\pm$ 2 \\
\hline
$\Delta t_{pulse}$ (ns)	& 5 &	\\
\hline
$A_{beam}$ (mm$^2$) 	& 23.15 &$\pm$ 0.43 \\
\hline
$\mathcal{T}_{window}$ 	& 0.801 &$\pm$ 0.023 \\
\hline
$I$ ($10^8$  W/cm$^2$) 	& 1.336 &$\pm$ 0.115	 
\end{tabular}
\end{table}

\end{comment}

%A strong wavelength dependence is expect, yet it was not observed in  \cite{Weinrotter-2004,Kopecek-Reider-2003,Tauer-2007,Ma-1998}. The electrons can be generated by the ionization of microparticles like dust and soot particles that absorb infrared radiation. Which agrees with the non-dependence on laser wavelength reported in 

%eed electrons originated by dust particles relate as well to the pressure dependence observed in our results, where lower pressure could not ignite without lens, but higher could. The decrease in pulse energy required to ignite a gas mixture with increasing pressure is reported by many authors in \cite{Kopecek-2003,Phuoc-White,Kopecek-Reider-2003,Phuoc-2000,Lee-2001}.

\section{Conclusion}
\label{Sec: Conclusion}

Laser-induced spark ignition is a process highly influenced by the gas filling pressure. The beam irradiance is key in determining the ignition success. We report that for pressure above 20 bar one can ignite a He--H$_2$--O$_2$ mixture with a non-focused laser. This translates to an irradiance around 10$^8$ W/cm$^2$, two to six orders of magnitude below the reported literature in Table \ref{Tab: MIE-Literature}. Despite being incapable of creating an macroscopic spark and visible electric breakdown in atmospheric air, the laser setup can still excite and ignite the mixture. The electrical breakdown may be initiated by a microscopic spark formed by impurities excited by the laser.

A laser absorption experiment was done to measure the energy absorption of a unfocused Nd:YAG laser at 1064~nm for two gas mixtures with pressures ranging between 10 and 100~bar. Less than 1\% (5\% with error bars) of the energy is absorbed by the gas \footnote{A more detailed error propagation analysis may be found in \cite{Ricardo-Thesis}, which means a very small portion of the laser pulse energy is sufficient for triggering ignition.} %In our setup this equates that 2~mJ are enough to start the combustion process. 
No significant differences in absorption were observed between the He--O$_2$ and He--H$_2$ mixtures, nor within the 10 to 100~bar of filling pressure range. 
The explanation for this non-focused laser ignition at high pressure may be related to the unintended presence of solid microparticles, like dust and soot, in the chamber. This explanation was already proposed by other authors \cite{Phuoc-2000, Phuoc-2006, Phuoc-White, Weinrotter-2004}, where impurities absorb the laser energy leading to high temperature spots, which produce free electrons. These will start the avalanche process previously described, by exciting the gas and  the chemical reactions. The initiation effect is expected to have no wavelength dependence, which is in agreement with the absence of electronic excitation transitions from ground state for He, O$_2$ or H$_2$ near 1064~nm. 

Further research is needed to more extensively understand why and in what conditions these events can take place. An experiment may be designed to measure minimum ignition energy and compare the focused and the unfocused regimes. Another possibility is to test ignition using ultra-high purity gases. The opposite option would be to seed the combustible mixture with dust/aerosol. Then the threshold pressures for unfocused ignition could be compared.

\begin{acknowledgements}
IPFN people received financial support from Funda\c{c}\~{a}o para a Ci\^{e}ncia e Tecnologia through projects UIDB/50010/2020 and UIDP/50010/2020 and the APPLAuSE PhD program through grant PD/BD/142970/2018. This work has been funded by the European Space Agency contract ESA/ESTEC 23086 ``Kinetic Shock-Tube for Planetary Exploration". The authors would like to acknowledge the assistance of Maria Castela in processing the combustion experiment pressure sensor data. The authors would also like to acknowledge the reviewers of this work for helping improve the quality of the manuscript.
\end{acknowledgements}

\bibliography{Brief-Communication-Biblio}

\end{document}